\documentclass[lnbip,sechang,a4paper]{svmultln}
\usepackage{amssymb}
\usepackage{graphicx}

\usepackage{url}
\urldef{\mailsa}\path|zheng.li@nicta.com.au|
\urldef{\mailsb}\path|xpllmmff@hotmail.com|
\urldef{\mailsc}\path|liamob99@hotmail.com|
\urldef{\mailsd}\path|dr.hezhang@gmail.com|   
\usepackage[pdfpagelabels,hypertexnames=false,breaklinks=true,bookmarksopen=true,bookmarksopenlevel=2]{hyperref}

\usepackage{array}
\usepackage{multirow}
\usepackage{enumitem}

\begin{document}

\mainmatter  

\title{The Cloud's Cloudy Moment: A Systematic Survey of Public Cloud Service Outage}

\titlerunning{A Systematic Survey of Public Cloud Service Outage}

%
%
\author{Zheng Li\inst{1,2} \and Mingfei Liang \inst{2} \and Liam O'Brien\inst{3} \and He Zhang\inst{4}}

\institute{CRL Lab, National Information and Communications Technology Australia (NICTA), 7 London Circuit, Canberra City, ACT 2601, Australia\\
\mailsa\\
\and Australian National University (ANU), Canberra, ACT 0200, Australia\\
\mailsb\\
\and ICT Innovation and Services, Geoscience Australia, Cnr Jerrabomberra Avenue and Hindmarsh Drive Symonston ACT 2609, Australia\\
\mailsc\\
\and State Key Laboratory of Novel Software Technology, Software Institute, Nanjing University, Jiangsu, China\\
\mailsd
}

%
%

\toctitle{Lecture Notes in Business Information Processing: Authors' Instructions}
\tocauthor{Authors' Instructions}
\maketitle

\begin{abstract}
Inadequate service availability is the top concern when employing Cloud computing. 
It has been recognized that zero downtime is impossible for large-scale Internet services. By learning from the previous and others' mistakes, nevertheless, it is possible for Cloud vendors to minimize the risk of future downtime or at least keep the downtime short. 
To facilitate summarizing lessons for Cloud providers, we performed a systematic survey of public Cloud service outage events. This paper reports the result of this survey.  
In addition to a set of findings, our work generated a lessons framework by classifying the outage root causes.  
The framework can in turn be used to arrange outage lessons for reference by Cloud providers. 
By including potentially new root causes, this lessons framework will be smoothly expanded in our future work.

\end{abstract}

\section{Introduction}
Cloud computing has increasingly become popular in the present business scenario, with various benefits ranging from convenience to economy. Many organizations are using Cloud to automate their service delivering. However, there are thorny issues and risks in using the Cloud \cite{Anthes_2010}. Among the numerous and different concerns \cite{Boampong_Wahsheh_2012}, \cite{Iankoulova_Daneva_2012}, \cite{Sun_Chang_2011}, the risk of inadequate service availability has been identified as the top obstacle to adoption of Cloud computing \cite{Armbrust_Fox_2010}, \cite{Sun_Chang_2011}. Given the reality that it is hard to eliminate the downtime of the data center systems behind Cloud services \cite{Bigelow_2011}, most of the current studies emphasized data/system backup from the perspective of Cloud customers. For example, employing multiple Cloud providers was suggested as being a plausible solution to very high availability service delivery \cite{Armbrust_Fox_2010}. 

Although there is no absolute means for preventing outage \cite{Bigelow_2011}, it is still worthwhile for Cloud vendors to learn from the existing mistakes, so as to minimize the risk of future downtime or at least keep the downtime short \cite{Pingdom_2008}. Unfortunately, to the best of our knowledge, few comprehensive investigations into service outages can be found in the literature. In other words, there is a lack of systematic discussion about outage lessons from the perspective of Cloud providers.

As an initial step to summarizing lessons for Cloud providers, we performed a survey of public Cloud service outages by using the Systematic Literature Review (SLR) approach. This paper reports the results of that survey together with the methodology of this systematic survey. Due to the limit of resource and time, our work only focused on the top five Cloud vendors. The collected outage data can then be viewed as the representative of all the existing Cloud service outage events. The corresponding data analysis was unfolded to answer four predefined research questions about the outage host, duration/frequency, location, and root cause.

The contribution of this work is twofold. First, based on the data analysis, we highlighted a set of findings (e.g.~two influential factors related to outage locations, cf.~Subsection \ref{RQ3}) when answering the predefined research questions. Second, by classifying the outage root causes, this study essentially generated a framework for accommodating outage lessons for Cloud providers.

The remainder of this paper is organized as follows. Section \ref{II} briefly introduces the methodology used to perform this survey. Section \ref{III} describes the survey result that answers the predefined research questions. Section \ref{lessonFramework} summarizes a set of sample lessons driven by the root cause classification. Conclusions and some future work are discussed in Section \ref{IV}.

\section{Methodology of this Survey}
\label{II}
Cloud service outage events have been generally reported as news or posts scattering over web media, technical websites, blogs, etc. To efficiently perform this survey, we borrowed SLR approach to collect, assess, and analyze the relevant outage reports. As the main methodology applied for Evidence-Based Software Engineering (EBSE) \cite{Dyba_Kitchenham_2005}, SLR has been widely accepted as a standard and systematic approach to investigation of specific research questions. Although the study objects here are not academic publications, this survey may still benefit from the rigorous review process defined by SLR. Following the guidelines of SLR \cite{Kitchenham_Charters_2007}, we did this work mainly covering three steps:
\begin{itemize}[leftmargin=30pt]
\renewcommand{\labelitemi}{$\bullet$}
    \item	Identify research questions and prepare the survey.
    \item	Collect relevant outage reports and extract data.
    \item	Analyze the extracted data and report the result. 
\end{itemize}

\subsection{Survey Preparation}
Similar to preparing an SLR, the preparation of this study generated a review protocol based on a pilot survey. Due to the limit of space, here we only highlight the research questions defined in the review protocol, as listed in Table \ref{tbl>1}.

\begin{table}[!t]\footnotesize
\renewcommand{\arraystretch}{1.3}
\centering
\caption{\label{tbl>1}Research Questions}
\begin{tabular}{l >{\raggedright}p{5cm}  >{\raggedright\arraybackslash}p{5cm}}
\hline

\hline
\textbf{ID~~~~} & \textbf{Research Question} & \textbf{Main Motivation}\\
\hline
RQ1 & Which Cloud provider experienced service outage? & To identify the outage host.\\
RQ2 & When and for how long did the service outage take place? & To identify the outage duration/frequency.\\
RQ3 & Where did the service outage happen? & To identify the outage location.\\
RQ4 & What is the root cause of the service outage? & To identify the outage reason.\\
\hline

\hline
\end{tabular}
\end{table}

\subsection{Report Collection and Data Extraction}
Unlike exploring various academic libraries in normal SLR, the outage report searching in this study only resorted to the Google search engine.

Furthermore, we employed three constraints for the outage report collection:
\begin{enumerate}[leftmargin=33pt]
\renewcommand{\labelenumi}{\it{(\theenumi)}}
    \item	This study only focused on public Cloud to make our effort closer to industry needs. Given the large number of players in the market \cite{Cloud_2013}, we further limited our concentration to a small set of top Cloud providers (cf.~Subsection \ref{providers}). 
    \item	Considering that the term ``Cloud computing" started to gain popularity in 2006 \cite{Zhang_Cheng_2010}, we only collected reports posted between 2007 and 2012. This study did not trace the old outage cases, though some Cloud services like Gmail or Hotmail have existed for a longer time.  
    \item	This study distinguished unplanned outages from all kinds of Cloud service downtime. In particular, we collected outage information only from the third-party media, so as to ignore the tiny issues that attracted little public attention.
\end{enumerate} 

In total, we collected 112 Cloud service outage events. By reading the outage details, we extracted useful data related to the pre-defined research questions for further analysis.\footnote{The extracted Cloud service outage data are shown online: \url{https://docs.google.com/spreadsheet/ccc?key=0AtKzcoAAmi43dEtPVVlIQ0NRblJiTV9SOGNJb2ttN0E}}

\subsection{Data Analysis}
The primary data analysis here is to carry out quantitative statistics based on the qualitative data descriptions. Given particular phenomena, we tried to give further explanations or suggestions. Moreover, the root causes of public Cloud service outage have been classified and arranged into a lessons framework for Cloud providers.

\section{Result of this Survey}
\label{III}
The survey results are organized and reported following the sequence of answers to the predefined research questions.

\subsection{RQ1: Which Cloud providers experienced service outage? }
\label{providers}

\begin{table}[!t]\footnotesize
\renewcommand{\arraystretch}{1.3}
\centering
\caption{\label{tbl>2}Top 5 Cloud Providers}
\begin{tabular}{c l c}
\hline

\hline
\textbf{Overall Rank~~~~} & \textbf{Cloud Provider~~~~} & \textbf{Number of Occurences}\\
\hline
1 & Amazon & 24 \\

2 & Rackspace& 19\\

3 & Microsoft & 18\\

4 & Google & 17\\

4 & Saleforce.com & 17\\
\hline

\hline

\end{tabular}
\end{table}

As mentioned previously, numerous public Cloud providers have been increasingly available in the market. It is thus nearly impossible to collect the outage data of different Cloud services all at once. Therefore, we decided to concentrate on the top Cloud providers only. Since different Cloud rankings have been published by different parties at different time, it would be more rational to combine those various opinions. By trying to exhaustively explore the web media and technical websites, we firstly gathered 34 rankings of public Cloud vendors.\footnote{The third-party rankings of public Cloud providers are listed online: \url{https://docs.google.com/spreadsheet/ccc?key=0AtKzcoAAmi43dE1TaTJINUdFM0hqVVQ3dy0wX0M3R2c}} Then, we rearranged the listed vendor names according to their occurrence numbers. As such, we finally achieved an overall Cloud ranking combining the individuals. Note that we deliberately excluded the Cloud rankings shown in the personal blogs. We found that most of the blogs were either criticized for bias or commented as copies of the others.

\begin{figure}[!t]
\centering
\includegraphics[width=10cm]{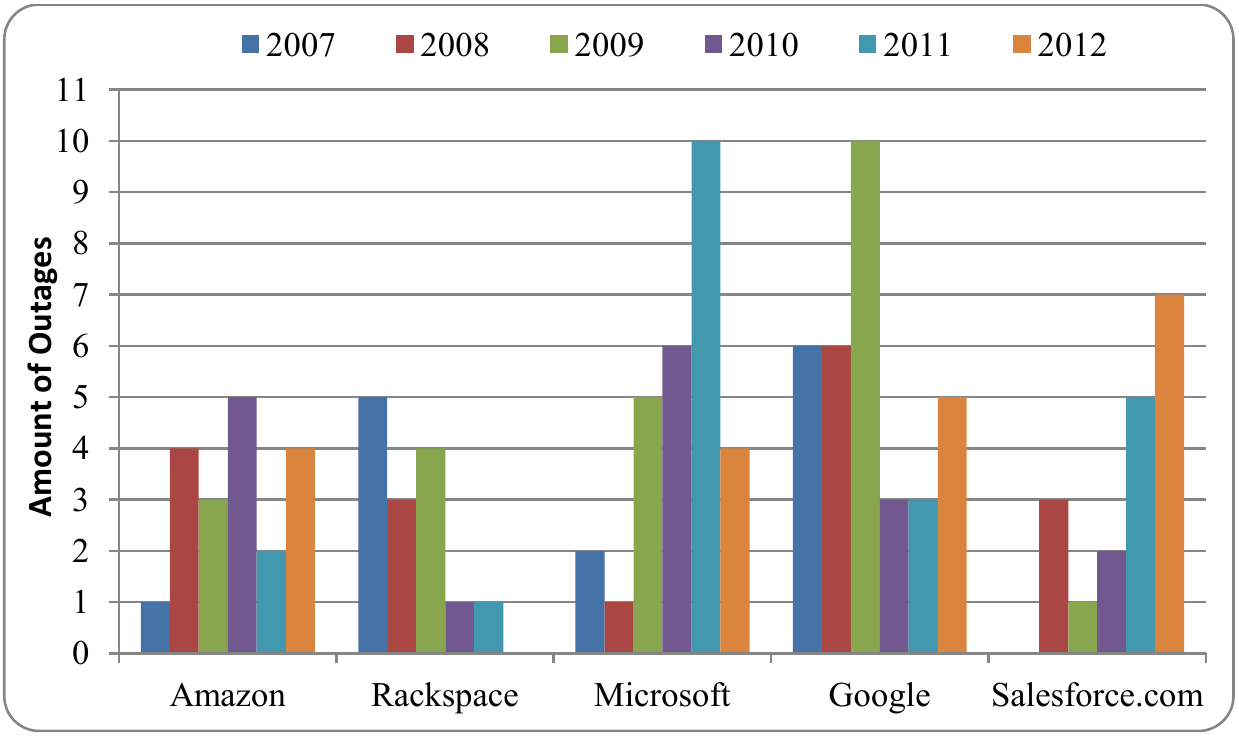}
\caption{Outage distribution over providers and years.}
\label{fig>PicProviderYear}
\end{figure}

Given the limited resource and time, we further narrowed our focus down to the top five public Cloud providers, as listed in Table \ref{tbl>2}. The collected data (cf.~Fig.~\ref{fig>PicProviderYear}) shows that each of the Cloud providers has suffered from considerable service outages, not to mention that there are also unreported downtime events. Such a phenomenon confirms the opinion that service outage happens to any Cloud provider sooner or later no matter how smart or successful the provider is \cite{Brooks_2010}, \cite{Pingdom_2008}. To reduce the risks of Cloud service outage, building redundancy could be a generic strategy for both Cloud vendors and customers \cite{Cogswell_2013}.

\subsection{RQ2: When and for how long did the service outage take place?}
\label{subsec>RQ2}
Through the outage distribution illustrated in Fig.~\ref{fig>PicProviderYear}, we also show that the top five Cloud providers suffer from service outages nearly every year. Two exceptions are: there was no outage report of Salesforce.com in 2007, and no report of Rackspace in 2012. The reason could be that Salesforce.com introduced its Platform-as-a-Service (PaaS) in late 2007 \cite{LaMonica_2007}, while Rackspace started expanding its old data center at the beginning of 2012 \cite{Miller_2012}.  

Furthermore, Cloud service outages could happen at any month. In particular, Amazon's northern Virginia data center seems more likely subjected to power outage during June (in 2008, 2009, and 2012), when thunderstorms start appearing frequently across Virginia every year \cite{Hayden_Michaels_2013}.

\begin{figure}[!t]
\centering
\includegraphics[width=10cm]{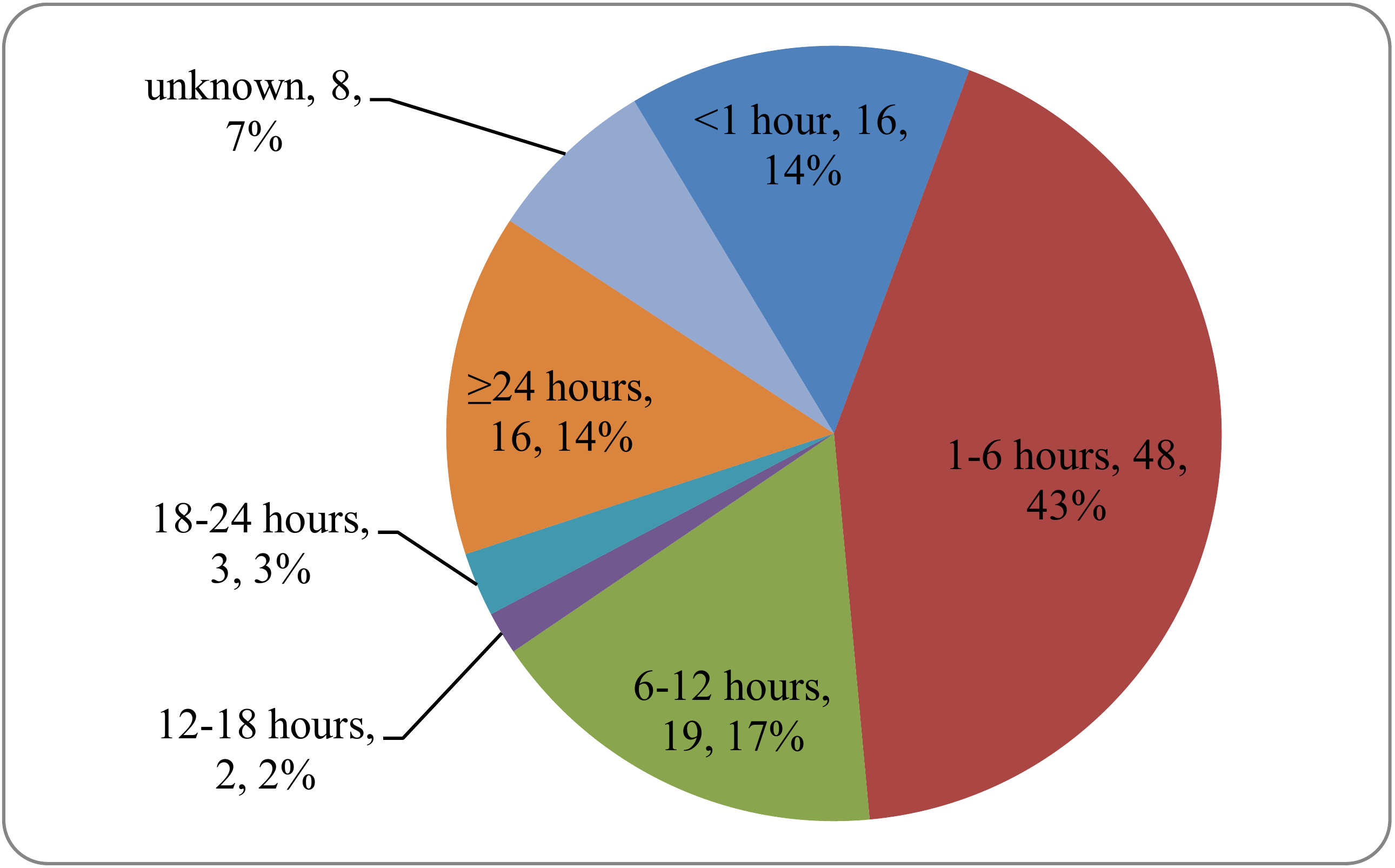}
\caption{Outage distribution over six-hour scales.}
\label{fig>PicDuration}
\end{figure}

As for the outage duration, we grouped the collected events into a set of six-hour scales, as illustrated in Fig.~\ref{fig>PicDuration}. Interestingly, the Cloud service outages lasted less than 12 hours and the others roughly follow the 80-20 distribution. In particular, we calculated several typical Cloud services' downtime and the corresponding availability, as shown in Table \ref{tbl>3}. It is clear that each Cloud provider has experienced violation of its Service Level Agreements (SLA) during particular service years. Note that the service downtime here refers to the sum of worst-case outage durations. For example, although Gmail only suffered an average of 10 to 15 minutes of downtime per month in 2008 \cite{BBC_2009}, an unlucky user could have lost the service for around 71.5 hours according to our collected data.

\begin{table}[!t]\footnotesize
\renewcommand{\arraystretch}{1.3}
\centering
\caption{\label{tbl>3}Typical Downtime of Cloud Services}
\begin{tabular}{l l l l}
\hline

\hline
\textbf{Cloud Service~~~~~~~~} & \textbf{Year~~~~} & \textbf{Downtime~~~~} & \textbf{Availability}\\
\hline
Amazon S3 & 2008 & $\sim$ 36.5 hours & $\sim$ 99.58\% \\

Amazon EC2 & 2011 & $\sim$ 8.5 days & $\sim$ 97.67\%\\

Rackspace Storage & 2011 & $>$ 48 hours & $<$ 99.45\%\\

Microsoft Azure & 2012 & $>$ 26.5 hours & $<$ 99.7\%\\

 Google Gmail & 2008 & $\sim$ 71.5 hours & $\sim$ 99.18\%\\

Saleforce.com Heroku & 2011 & $\sim$ 104 hours & $\sim$ 98.81\%\\

\hline

\hline
\end{tabular}
\end{table}

Therefore, we suggest that the industry should help educate Cloud consumers to ``expect the unexpected" before using outages. It is understandable that Cloud providers tend to particularly emphasize their SLA for the purpose of market hype. However, the ideal claims could mislead customers and in turn spoil the Cloud ecosphere. For example, users have considered that Cloud ``downtime is completely unacceptable" \cite{McLaughlin_2008}, while the truth is that there is no absolute means for preventing downtime when running large-scale Internet services \cite{Bigelow_2011}.

\subsection{RQ3: Where did the service outage happen?}
\label{RQ3}
When extracting data, we found that a large proportion of outage events did not disclose their geographical locations, especially the Software-as-a-Service (SaaS) ones. Only Amazon- and Rackspace-related outage reports mostly specified the data centers where the service outages happened. Therefore, we mainly focused on Amazon and Rackspace to answer this research question. Given the reports specifying locations, 72.2\% of Amazon outages (13/18) took place in its northern Virginia data center, while 72.7\% of Rackspace outages (8/11) happened in its Dallas-Fort Worth (DFW) data center. By roughly investigating these two places, we summarized two influential factors related to the locations of Cloud data centers, namely \textit{climate} and \textit{time}.

\textit{Climate Influence:} As mentioned previously, power outages occurred with thunderstorms three times (once a year) in Amazon's northern Virginia data center. In fact, thunderstorms are a frequent concern in Virginia, although northern Virginia experiences the least number of such storms \cite{Hayden_Michaels_2013}. Recall that ``away from natural disasters" is one of the principles of site selection for building a data center \cite{Fontecchio_2007}, Amazon may have put this data center at risk from the beginning.

\textit{Time Influence:} As one of the oldest Rackspace data centers, DFW data center has experienced a series of equipment failures \cite{Miller_2012}. Interestingly, the northern Virginia data center is also one of the Amazon's oldest. We are then concerned with two points for this phenomenon: on the one hand, old data centers may involve immature techniques and mechanisms from the beginning; on the other hand, a data center could gradually become vulnerable with equipment aging as time goes by. As such, a natural suggestion is that the Cloud data centers should be upgraded regularly.

\begin{table}[p]\footnotesize
\renewcommand{\arraystretch}{1.2}
\centering
\caption{\label{tbl>4}Outage Root Cause Classification}
\begin{tabular}{l l l}
\hline

\hline
\multicolumn{3}{c}{\textbf{Root Cause of Public Cloud Service Outage}}\\
\hline
& \multicolumn{2}{l}{Direct Power Cut/Interruption}\\
\cline{2-3}
\multirow{13}{*}{Power Outage~~~~} & \multirow{8}{*}{Hardware~~~~} & Breaker \\

& & Bus Duct\\

& & Cable\\

& & Electrical Ground\\

& & Power Distribution Unit (PDU)\\

& & Programmable Logic Controllers\\

& & Transfer Switch\\

& & Utility Distribution Network\\
\cline{2-3}

& \multicolumn{2}{l}{Human Mistake}\\
\cline{2-3}
& \multicolumn{2}{l}{Natural Disaster}\\
\cline{2-3}

& \multicolumn{2}{l}{Uninterruptible Power Supply (UPS) Issue}\\
\cline{2-3}
& \multicolumn{2}{l}{Vehicle Accident}\\
\hline
\multirow{10}{*}{Routing/Network Issue~~~~} & \multicolumn{2}{l}{DNS Error} \\
\cline{2-3}
& \multirow{3}{*}{Hardware} & Core Device \\

& & Infrastructure\\

& & Routing Device\\
\cline{2-3}
& \multirow{2}{*}{Human Mistake~~~~} & Misconfiguration \\

& & Misoperation\\
\cline{2-3}
& \multicolumn{2}{l}{Request Flood} \\
\cline{2-3}
& \multirow{3}{*}{Software} & Bug \\

& & Communication Error\\

& & HTTP Error\\
\hline
\multirow{13}{*}{(Other) System Issue~~~~} & \multicolumn{2}{l}{Database Error} \\
\cline{2-3}
& \multirow{2}{*}{Hack} & DDoS Attack \\

& & Virus\\
\cline{2-3}
& \multirow{3}{*}{Hardware} & Chiller Failure \\

& & Recent Change\\

& & Server Down\\
\cline{2-3}
& \multirow{2}{*}{Human Mistake} & Misconfiguration \\

& & Misoperation\\
\cline{2-3}
& \multirow{2}{*}{Overload} & Memory Leak \\

& & Request Flood\\
\cline{2-3}
& \multirow{2}{*}{Software} & Bug \\

& & Recent Change\\
\cline{2-3}
& \multicolumn{2}{l}{Storage Error} \\
\hline
\multicolumn{3}{l}{Third-party Outage}\\
\hline

\hline
\end{tabular}
\end{table}

\subsection{RQ4: What is the root cause of the service outage?}

\begin{figure}[!t]
\centering
\includegraphics[width=8.5cm]{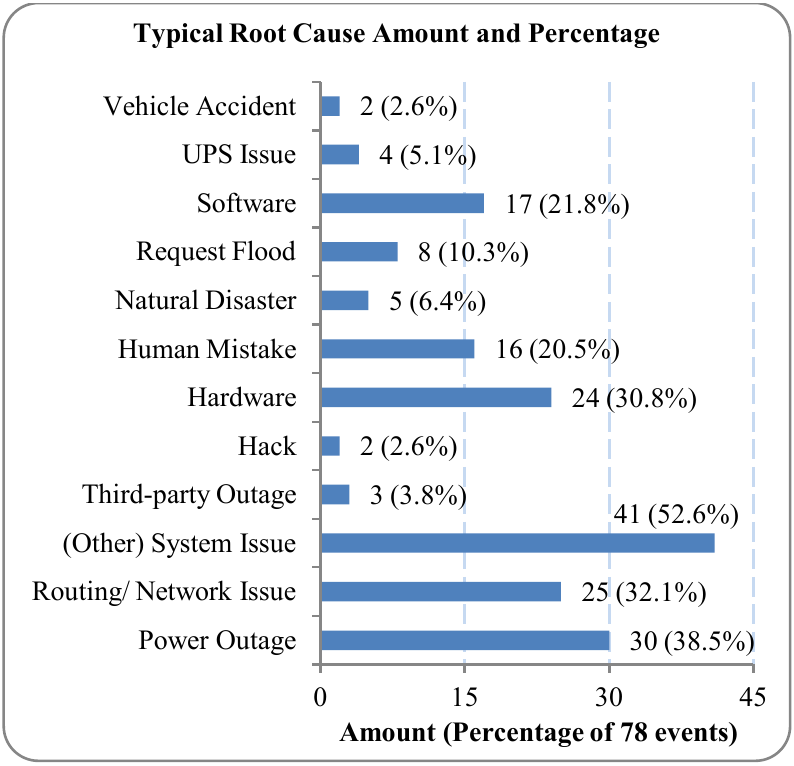}
\caption{Distribution of several typical root causes. Note that, for the convenience of comparison between root causes, we summed up the outage events with the same sub-category name, although they could be under different primary cause classes.}
\label{fig>PicOutageDistribution}
\end{figure}

Given the collected reports, it is impossible to identify the root cause of every Cloud service outage event. 
For example, Cloud providers may decline to supply technical details (e.g.~Google News outage on September 22, 2009). Therefore, we only focused on the 78 out of 112 events with outage cause explanations. In addition, since an outage event may be a result of a combination of causes (e.g.,~Microsoft Office365 outage on November 13, 2012) or a close cause chain (e.g.,~Amazon EBS outage on October 22, 2012), we further broke the 78 outage explanations into 99 cause units.\footnote{The breakdown of the outage causes are listed online: \url{https://docs.google.com/spreadsheet/ccc?key=0AtKzcoAAmi43dDdCUjhuTlZiaXJXWEZNQ1FGTTB2blE}} By classifying those units (cf.~Table \ref{tbl>4}), we show a set of typical and relatively frequent root causes of Cloud service outages, as illustrated in Fig.~\ref{fig>PicOutageDistribution}.

In general, \textit{Power Outage} and \textit{Routing/Network Issue} are two common classes of Cloud service outage causes. The large amount of routing/network issues may not be surprising because the Clouds are inherently associated with intranet and Internet, but the power supply behind Cloud services seems more vulnerable than we expected. As for the details of power outage, it can be confirmed that Uninterruptible Power Supply (UPS) is not uninterruptible enough \cite{McFarlane_2005}; thunderstorm (lightning strike) is currently the only natural threat to power equipments; while interestingly, vehicle accident is not rare for being a reason of power interruption, and we thus suggest that the physical barrier security rule \cite{Higbie_2005} should be applied not only to the Cloud data center buildings but also to the outside power infrastructures.

The other three common cause sub-categories are \textit{Hardware}, \textit{Software}, and \textit{Human Mistake}. Each of the three categories covers more than one fifth outage events. In particular, the \textit{Third-party Outage} refers to the scenario that an outage event happens due to other Cloud service outages (e.g.,~Heroku outage on July 10, 2012). This cause type suggests that not only Cloud consumers but also providers may suffer from the ``Cloud ripple effect" \cite{Darrow_2012}.

Overall, the root cause classification shown in Table \ref{tbl>4} can be viewed as a lessons framework for Cloud providers (cf.~Section \ref{lessonFramework}). The large amount of lessons learnt from the existing Cloud service outage events can be naturally organized by using this framework. Through smooth expansion, we may continually develop this framework to cover new type of Cloud service outages and accommodate new lessons. 



\section{Sample Lessons from Cloud Service Outage}
\label{lessonFramework}
Driven by the outage root cause classification (cf.~Table \ref{tbl>4}), we have tried to summarize and rationalize the existing lessons from the Cloud service outage events. Due to the limit of space, here we only show some coarse-grained sample lessons that can be matched up to those primary cause classes. 

\subsection{Watch the Power}
It has been claimed that ``the worst, most sustained downtime has always been caused by power issues", while losing power in data centers will happen someday inevitably \cite{Kaplan_Moss_2009}. Therefore, we elaborate relatively more on the power-related lessons as follows.

\subsubsection{Any single piece of power equipments can fail.} The failed equipment piece could incur cascading events and result in a large scale of power outage. For example, a failed bus duct prevented proper operation of Uninterruptible Power Supply (UPS) (cf.~Rackspace outage on July 7, 2009); an electrical ground fault and short circuit in a major power distribution panel interrupted power to a particular Availability Zone (cf.~Amazon outage on May 8, 2010); the failure of switches in the electrical infrastructure prevented transfer of electrical load between different power sources (cf.~Rackspace outage on June 29, 2009); a breaker failure in the switch board affected all downstream power distribution units (PDUs) (cf.~Rackspace outage on November 11, 2007); while failures in a PDU resulted in a portion of servers losing power (cf.~Amazon outage on December 9, 2009).

Moreover, some power equipment failures could happen externally. For example, there could be unexpected power cut (e.g.,~Rackspace outage on December 5, 2007); and even small problems with utility distribution system (e.g.,~Amazon outage on June 14, 2012) can cause power outage. 

\subsubsection{Uninterruptible Power Supply (UPS) is not uninterruptible enough.} Given the possibly vulnerable power equipment, employing redundant/backup power systems would be a natural strategy to reduce power issues. An ideal mechanism could be ``multiple power supplies in every server connected to 2 PDUs connected to 2 different generators" \cite{Hacker_2012}. Considering the cost-benefit tradeoffs, one of the most practical and common efforts is to use UPS. Unfortunately, this study shows that UPSes could also become a huge single point of failure. Interestingly, UPS has been criticized for its deceptive designation: Once out of commission, UPS can be a solid barrier between Cloud service and generator power \cite{McFarlane_2005}. Therefore, regular testing should be further emphasized even for ``uninterruptible" power equipments. 

\subsubsection{Backup power systems should be tested regularly.} Recall the Amazon outage on June 14, 2012, even with the correct setup of generator fallback, the power backup mechanism could still fail unexpectedly in some circumstance. It has been pointed out that the lack of regular testing is the backend flaw, although the power interruption could be the head of a cause chain \cite{Whittaker_2012}. In fact, regularly testing the backup power systems has been strongly suggested by industry \cite{Kaplan_Moss_2009}.

\subsection{Be Pessimistic about every Service Component}
Given the listed root causes of public Cloud service outage (cf.~Table \ref{tbl>4}), it is clear that various hardware and software issues can knock out Cloud services, not to mention the numerous routing/network problems (cf.~Fig.~\ref{fig>PicOutageDistribution}). As such, it would be valuable and necessary for people to realize and understand that ``Clouds are made of components that can fail" \cite{Hammar_2012}. As mentioned in Section \ref{subsec>RQ2}, Cloud consumers should be ready to ``expect the unexpected" outages, while Cloud providers should rethink and carefully build service levels that actually guarantee services from the perspective of consumers \cite{Karstens_2011}. Note that understanding such a reality does not mean to ask people to passively live with it. On the contrary, being pessimistic about every service component requires both Cloud providers and consumers to fine tune their processes and responses to failures, by conducting full-blown load tests of their failover mechanisms \cite{Hammar_2012}, \cite{Rasmussen_2012}.

\subsection{Minimize the Chain Reaction}
As previously mentioned, we find that an event of Cloud service outage could often comprise a combination of causes or a close cause chain. Some discussions revealed that ``the stress of failure will trigger a cascade of other failures" \cite{Hammar_2012}. One of the logics behind this advice could be that human beings tend to make more mistakes under pressure. Inspired by the fire drills that help train people to deal with the event of an emergency, frequent load tests of failover plans may help engineers get familiar with what they need to do to reduce human mistakes when fixing Cloud service outages.

When it comes to the ``Cloud ripple effect" \cite{Darrow_2012} resulted from third-party outages, we may draw similar lessons for both Cloud service providers and consumers. In fact, the secondary or tertiary Cloud service providers are indeed customers of their primary providers. Interestingly, a consensus on surviving third-party outages is all about redundancy. Following the terminology from Amazon, the suggestion is to spread the load across multiple availability zones and across multiple regions \cite{Butler_2012}, \cite{Dave_2013}; in general, the suggestion is to spread across multiple data centers and across multiple primary providers \cite{Butler_2012}, \cite{Dave_2013}; a more aggressive suggestion is even to spread across public and private Clouds \cite{Horovits_2012}. There is little doubt that such load spreading could be complicated and expensive, however, it would be a worthwhile mechanism if the Cloud services are serious about customer satisfaction \cite{Paterson_2013}.

\subsection{Open the Outage Details}
According to the collected outage data, we find that many events of public Cloud service outage did not disclose the details. It is natural that the ``public Cloud" implies some loss of control and visibility from the customers' perspective. Nevertheless, delivering enough in-depth information has been identified crucial for customers especially during the outage \cite{Karstens_2011}, \cite{Hickey_2011}, \cite{Ainsworth_2011}. More importantly, opening outage details would also be beneficial for Cloud service providers. There are two reasons for this. 

Firstly, rapid and clear lines of communication have been proved a successful crisis management model. The existing case studies show that keeping customers updated can significantly drop off negative commentary \cite{Ainsworth_2011}. Given the timely disclosure of an outage and the remedy activities thereafter, most customers would still forgive the Cloud service providers for their failings \cite{Wainewright_2011}.

Secondly, disclosure of outage details can help boost the entire Cloud computing industry. A positive observation on Cloud service outages is that those unfortunate events also provided opportunities to learn from them \cite{Rasmussen_2012}. By exposing what went wrong, each outage essentially acts as an education for Cloud service providers with how to prevent it from happening again or how to adapt when an outage occurs. For example, the existing Cloud failures have provided useful lessons in disaster planning and infrastructural designing for redundancy, which would reduce future risks and eventually make the Cloud stronger \cite{Pingdom_2008}, \cite{Hickey_2011b}.

\section{Conclusions and Future Work}
\label{IV}
There is no doubt that an outage of Cloud services happens sooner or later. As such, it is necessary and worthwhile for Cloud providers to learn from the previous and others' mistakes to minimize the risks of future downtime \cite{Pingdom_2008}. This paper reports a systematic survey of public Cloud service outages. In addition to revealing findings based on the quantitative analysis, this study finally establishes an education framework for learning from Cloud service outage. We also list a set of coarse-grained sample lessons to show that the established framework can help guide people to arrange and/or refer to the relevant knowledge.

The main limitation of this work is the completeness of the Cloud service outage data. On the one hand, we only focused on a small amount of public Cloud vendors. On the other hand, we only collected outage events reported by web media and technical websites. Therefore, our future work will further collect outage events of more Cloud vendors and gradually expand the lessons framework. Meanwhile, we will continue summarizing the outage lessons 
and arrange them within the aforementioned framework for reference by Cloud providers. 
\section*{Acknowledgments}

NICTA is funded by the Australian Government as represented by the Department of Broadband, Communications and the Digital Economy and the Australian Research Council through the ICT Centre of Excellence program.

\footnotesize \bibliographystyle{splncs}
\bibliography{ref}

\begin{thebibliography}{10}

\bibitem{Anthes_2010}
Anthes, G.:
\newblock Security in the {C}loud.
\newblock Commun. ACM \textbf{53}(11) (November 2010)  16--18

\bibitem{Boampong_Wahsheh_2012}
Boampong, P.A., Wahsheh, L.A.:
\newblock Different facets of security in the {C}loud.
\newblock In: Proc.~15th Communications and Networking Simulation Symp.~(CNS
  2012), Orlando, FL, USA, Society for Computer Simulation International (March
  26 - 29 2012)  1--7

\bibitem{Iankoulova_Daneva_2012}
Iankoulova, I., Daneva, M.:
\newblock Cloud computing security requirements: {A} systematic review.
\newblock In: Proc.~6th Int.~Conf.~Research Challenges in Information Science
  (RCIS 2012), Valencia, Spain, IEEE Computer Society (May 16-18 2012)  1--7

\bibitem{Sun_Chang_2011}
Sun, D., Chang, G., Sun, L., Wang, X.:
\newblock Surveying and analyzing security, privacy and trust issues in {Cloud}
  computing environments.
\newblock Procedia Eng. \textbf{15} (2011)  2852--2856

\bibitem{Armbrust_Fox_2010}
Armbrust, M., Fox, A., Griffith, R., Joseph, A.D., Katz, R., Konwinski, A.,
  Lee, G., Patterson, D., Rabkin, A., Stoica, I., Zaharia, M.:
\newblock A view of {C}loud computing.
\newblock Commun. ACM \textbf{53}(4) (April 2010)  50--58

\bibitem{Bigelow_2011}
Bigelow, S.:
\newblock The causes and costs of data center system downtime: {A}dvisory board
  {Q\&A}.
\newblock
  \url{http://searchdatacenter.techtarget.com/feature/The-causes-and-costs-of-%
data-center-system-downtime-Advisory-Board-QA} (June 2011)

\bibitem{Pingdom_2008}
Pingdom:
\newblock The major internet outages so far in 2008.
\newblock
  \url{http://royal.pingdom.com/2008/09/04/the-major-internet-outages-so-far-i%
n-2008/} (September 2008)

\bibitem{Dyba_Kitchenham_2005}
Dyb{\aa}, T., Kitchenham, B.A., J{\o}rgensen, M.:
\newblock Evidence-based software engineering for practitioners.
\newblock IEEE Softw. \textbf{22}(1) (January 2005)  58--65

\bibitem{Kitchenham_Charters_2007}
Kitchenham, B.A., Charters, S.:
\newblock Guidelines for performing systematic literature reviews in software
  engineering.
\newblock Technical Report EBSE 2007-001, Keele Univ. and Durham Univ. Joint
  Rep. (2007)

\bibitem{Cloud_2013}
CloudHarmony:
\newblock {Public Clouds}.
\newblock \url{http://cloudharmony.com/clouds} (Feburary 2013)

\bibitem{Zhang_Cheng_2010}
Zhang, Q., Cheng, L., Boutaba, R.:
\newblock Cloud computing: {S}tate-of-the-art and research challenges.
\newblock J. Internet Serv. Appl. \textbf{1}(1) (May 2010)  7--18

\bibitem{Brooks_2010}
Brooks, C.:
\newblock Heroku learns from {Amazon EC2} outage.
\newblock
  \url{http://searchcloudcomputing.techtarget.com/news/1378426/Heroku-learns-f%
rom-Amazon-EC2-outage} (January 2010)

\bibitem{Cogswell_2013}
Cogswell, J.:
\newblock Building redundancy into your {Cloud} outage strategy.
\newblock
  \url{http://searchcloudcomputing.techtarget.com/tip/Building-redundancy-into%
-your-cloud-outage-strategy} (January 2013)

\bibitem{LaMonica_2007}
LaMonica, M.:
\newblock Salesforce.com extends its application platform with {Force.com}.
\newblock \url{http://news.cnet.com/8301-10784_3-9778674-7.html} (September
  2007)

\bibitem{Miller_2012}
Miller, R.:
\newblock Rackspace {Cloud} will expand in {Dallas}.
\newblock
  \url{http://www.datacenterknowledge.com/archives/2012/01/06/rackspace-cloud-%
will-expand-in-dallas/} (January 2012)

\bibitem{Hayden_Michaels_2013}
Hayden, B.P., Michaels, P.J.:
\newblock Virginia's climate.
\newblock \url{http://climate.virginia.edu/description.htm} (February 2013)

\bibitem{BBC_2009}
BBC:
\newblock Gmail down again for some users.
\newblock \url{http://news.bbc.co.uk/2/hi/technology/7934443.stm} (March 2009)

\bibitem{McLaughlin_2008}
McLaughlin, K.:
\newblock Microsoft {Windows Live} services suffer global outage.
\newblock
  \url{http://www.crn.com/news/applications-os/206900295/microsoft-windows-liv%
e-services-suffer-global-outage.htm} (February 2008)

\bibitem{Fontecchio_2007}
Fontecchio, M.:
\newblock Study ranks cheapest places to build a data center.
\newblock
  \url{http://searchdatacenter.techtarget.com/news/1238054/Study-ranks-cheapes%
t-places-to-build-a-data-center} (January 2007)

\bibitem{McFarlane_2005}
McFarlane, R.:
\newblock {UPS} -- it's {NOT} uninterruptible.
\newblock
  \url{http://searchdatacenter.techtarget.com/news/1148907/UPS-its-NOT-uninter%
ruptible} (November 2005)

\bibitem{Higbie_2005}
Higbie, C.:
\newblock Rules for designing the urban data center.
\newblock
  \url{http://searchdatacenter.techtarget.com/tip/Rules-for-designing-the-urba%
n-data-center} (April 2005)

\bibitem{Darrow_2012}
Darrow, B.:
\newblock Heroku stung by amazon outage.
\newblock \url{http://gigaom.com/2012/06/15/heroku-stung-by-amazon-outage/}
  (June 2012)

\bibitem{Kaplan_Moss_2009}
Kaplan-Moss, J.:
\newblock Lessons from rackspace's downtime.
\newblock \url{http://jacobian.org/writing/lessons-from-rackspace-downtime/}
  (November 2009)

\bibitem{Hacker_2012}
{Hacker News}:
\newblock {Cascading errors caused AWS to go down}.
\newblock \url{https://news.ycombinator.com/item?id=4124719} (2012)

\bibitem{Whittaker_2012}
Whittaker, Z.:
\newblock Amazon explains latest cloud outage: Blame the power.
\newblock
  \url{http://www.zdnet.com/blog/btl/amazon-explains-latest-cloud-outage-blame%
-the-power/80094} (June 2012)

\bibitem{Hammar_2012}
Hammar, S.:
\newblock Lessons learned from the amazon web services outage.
\newblock
  \url{http://blog.apicasystem.com/2012/10/24/lessons-learned-from-the-amazon-%
web-services-outage/} (October 2012)

\bibitem{Karstens_2011}
Karstens, A.:
\newblock Lessons from the amazon outage: 5 ways that cloud providers must take
  responsibility for service levels.
\newblock
  \url{http://blogs.ixiacom.com/ixia-blog/amazon-outage-cloud-provider-service%
-levels/} (May 2011)

\bibitem{Rasmussen_2012}
Rasmussen, A.:
\newblock Lessons learned from cloud outages.
\newblock \url{http://apmdigest.com/lessons-learned-from-cloud-outages} (July
  2012)

\bibitem{Butler_2012}
Butler, B.:
\newblock 5 tips for surviving a cloud outage.
\newblock
  \url{http://www.networkworld.com/news/2012/042712-cloud-outage-tips-258736.h%
tml} (April 2012)

\bibitem{Dave_2013}
Dave, A.:
\newblock Learn to fail and avoid the next cloud outage.
\newblock
  \url{http://www.networkworld.com/news/tech/2013/021113-cloud-outage-266604.h%
tml?page=1} (February 2013)

\bibitem{Horovits_2012}
Horovits, D.:
\newblock {AWS} outage - moving from multi-availability-zone to multi-cloud.
\newblock
  \url{http://www.cloudifysource.org/2012/10/24/aws-outage-multi-availability-%
zone-multi-cloud.html} (October 2012)

\bibitem{Paterson_2013}
Paterson, J.:
\newblock Amazon {EC2} outages - lessons not learned.
\newblock
  \url{http://www.channelweb.co.uk/crn-uk/view-from-the-channel-blog/2292069/a%
mazon-ec2-outages-lessons-not-learned} (September 2013)

\bibitem{Hickey_2011}
Hickey, A.R.:
\newblock Amazon cloud outage highlights need for transparency.
\newblock
  \url{http://www.crn.com/news/cloud/229402233/amazon-cloud-outage-highlights-%
need-for-transparency.htm} (April 2011a)

\bibitem{Ainsworth_2011}
Ainsworth, J.:
\newblock Outages: Cloud customers cry out for communication.
\newblock
  \url{http://smartdatacollective.com/jamesainsworthalteriancom/39151/cloud-cu%
stomers-cry-out-communication} (August 2011)

\bibitem{Wainewright_2011}
Wainewright, P.:
\newblock Seven lessons to learn from amazon's outage.
\newblock
  \url{http://www.zdnet.com/blog/saas/seven-lessons-to-learn-from-amazons-outa%
ge/1296} (April 2011)

\bibitem{Hickey_2011b}
Hickey, A.R.:
\newblock Amazon's outage will make the cloud stronger.
\newblock
  \url{http://www.crn.com/slide-shows/cloud/229402271/amazon-cloud-outage-10-l%
essons-learned.htm?pgno=9} (April 2011b)

\end{thebibliography}

\end{document}